\documentclass[aps]{revtex4}

\usepackage{graphicx}%
\usepackage{dcolumn}%
\usepackage{bm}%

\usepackage{amssymb}
\usepackage{amsmath}

\begin{document}

\title{Mathematical treatment of the canonical finite state machine for the Ising model: $\epsilon$-machine}

\author{E. \surname{Rodr\'iguez-Horta}}
\affiliation{Facultad de F\'isica-Instituto de Ciencias y Tecnolog\'ia de Materiales, University of Havana (IMRE), San Lazaro y L. CP 10400. La Habana. Cuba.}

\author{E. \surname{Estevez-Rams}}
\email{estevez@imre.oc.uh.cu}
\affiliation{Facultad de F\'isica-Instituto de Ciencias y Tecnolog\'ia de Materiales, University of Havana (IMRE), San Lazaro y L. CP 10400. La Habana. Cuba.}

\author{R. \surname{Lora Serrano}}
\affiliation{Universidade Federal de Uberlandia, AV. Joao Naves de Avila, 2121- Campus Santa Monica, CEP 38408-144, Minas Gerais, Brasil. }

\begin{abstract}
The complete framework for the $\epsilon$-machine construction of the one dimensional Ising model is presented correcting previous mistakes on the subject. The approach follows the known treatment of the Ising model as a Markov random field,  where usually the local characteristic are obtained from the stochastic matrix, the problem at hand needs the inverse relation, or how to obtain the stochastic matrix from the local characteristics, which are given via the transfer matrix treatment. The obtained expressions allow to perform complexity-entropy analysis of particular instance of the Ising model. Three examples are discussed: the 1/2-spin nearest neighbor and next nearest neighbor Ising model, and the persistent biased random walk.
\end{abstract}

\date{\today}

\maketitle

\section{Introduction}

$\epsilon$-machine reconstruction or computational mechanics is an approach to discover the nature of patterns and to quantify it \cite{crutchfield92,crutchfield12}. Building from information theory concepts, it has found applications in several fields and proved its value in a number of context\cite{varn13,ryabov11,haslinger10}. Its use in statistical mechanics allows to define and calculate magnitudes that complements thermodynamic quantities such as entropy, specific heat, correlation  length or structure factor\cite{feldman98,feldman98a}. For a stochastic process considered to be stationary, the $\epsilon$-machine is its optimal minimal description, understood, as having the best (most accurate) predictive power while using the least possible resources (minimal statistical complexity)\cite{crutchfield94}. In computational mechanics, the notion of causality is taken in a general temporal sense, in a given context, cause to effect relations are established between past to future events\cite{shalizi01}.

Although a general framework, computational mechanics has been thoroughly developed for discrete time and space process by several authors \cite{varn13,shalizi01,lohr09,lohr09a}.

One of the earlier developments in computational mechanics has been the analysis of patterns in one dimensional spin systems under Ising-type models\cite{feldman98,feldman98a,crutchfield97,feldman08}. The use of the $\epsilon$-machine reconstruction for the one dimensional Ising model allowed the analysis of such systems as a deterministic finite state machine, and close expressions for the entropy density, statistical complexity and excess entropy were found\cite{feldman98,feldman98a,crutchfield97}. Unfortunately, this treatments failed to understand that the Ising model has to be considered a Markov (Gibbs) random field in order to correctly determine the probability measure of the associated Markov process\cite{behrends00}. The failure of not considering correctly the Markov random field, leads to incorrect expressions for the stochastic matrix governing the conditional probability between spin blocks. As a consequence, all quantitative results derived from such matrix are incorrect and this includes the determination of the causal states, the statistical complexity, the entropy density and the excess entropy. This includes the worked example on the Ising 1/2-nearest neighbor model and next nearest neighbor model.  

In this contribution the framework of computational mechanics for the Ising model from the most general setting in one dimension is developed. It is shown how to obtain the stochastic matrix from the Markov field: an inverse problem  not usually treated on the literature.  The results goes beyond correcting the mathematical treatment. An important consequence of the presented mathematical development is whether is it possible to cast the Ising model, in the general case, a single emission spin sequential process. We finally show the use of the developed framework via three examples.

The paper is organized as follows. In section \ref{sec:intr} the transfer matrix method for the Ising model is shortly reviewed for self-completeness and to fix the used notation. In section \ref{sec:randomfield} we describe the Ising model as a random field and deduce the expression for the stochastic conditional matrix. In section \ref{sec:epsilonmachine}, building from the previous sections, the $\epsilon$-machine formalism for the Ising model is discussed and the entropic magnitudes are introduced. Finally, in section \ref{sec:example} worked examples applying the developed mathematical tools is presented. Conclusions follow. 

\section{The transfer matrix formalism}\label{sec:intr}

The Ising (-Lenz) model is probably the most studied lattice type model in statistical mechanics, and is well covered in several statistical physics books for nearest 1/2 spin neighbor interactions\cite{lavis99}. Let us briefly recap, for completeness and notation purposes, the basic ideas of the transfer matrix formalism (we closely follow Dobson\cite{dobson69}) but in a general setting of a local type interaction Hamiltonian which is usually not found in texts. 

Consider an one dimensional chain of discrete values of length $L$:
\[ s^{L}=s_{0}s_{1}s_{2}\ldots s_{L-1}, \]
where $s_{i}$ can take values from a finite alphabet $\Theta$ of cardinality $\theta(=|\Theta|)$ (there will be $\theta^{L}$ possible sequences $s^{L}$). Each individual $s_{i}$ is called a spin. The interaction energy between spins of the sequence has a finite range $n$ such that it can be written as
\begin{equation}
 \displaystyle E(s_{i}, s_{i\pm k})= \left \{ \begin{array}{rr}\Lambda(s_{i},s_{i\pm k}) & 0 < k \leq n \\ 0 & k > n\end{array}\right. .\label{finiteenergy}
\end{equation}

The $s^{L}$ sequence can be partitioned into blocks of length $n$, considering $L=Nn$,
\[ s^{L}=[s_{0}s_{1}\ldots s_{n-1}][s_{n}s_{n+1}\ldots s_{2n-1}]\ldots [s_{(N-1)n}s_{(N-1)n+1}\ldots s_{Nn-1}], \]
which can be relabeled
\[ \begin{array}{ll}s^{L}&=[s^{(0)}_{0}s^{(0)}_{1}\ldots s^{(0)}_{n-1}][s^{(1)}_{0}s^{(1)}_{1}\ldots s^{(1)}_{n-1}]\ldots [s^{(N-1)}_{0}s^{(N-1)}_{1}\ldots s^{(N-1)}_{n-1}].  \\ &=\eta_{0}\eta_{1}\ldots \eta_{N-1}.\end{array}\]
Where
\begin{equation}
  \eta_{i}=s^{(i)}_{0}s^{(i)}_{1}\ldots s^{(i)}_{n-1}. \label{eta}
\end{equation}
The set of all possible blocks $\eta_{i}$ will be denoted by $\Upsilon$ with cardinality $\upsilon=\theta^{n}$. $\Upsilon$ will be taken as an ordered set (e.g lexicographic order) and to each $\eta_{i}$ a natural number, between $0$ and $\upsilon-1$, will be assigned. In what follows, $\eta_{i}$ should be understood not only as the configuration (\ref{eta}) but also as its corresponding order in the set $\Upsilon$, context will eliminate any ambiguity.

As the interaction energy has range $n$, one spin corresponding to the $\eta_{i}$ block, can only interact with all the spins within $\eta_{i}$ (type I) and at least one spin from the adjacent blocks $\eta_{i\pm 1}$ (type II).  

Taking into account that  $\Lambda(s_{i},s_{j})=\Lambda(s_{j},s_{i})$, the interaction energy of type I for the $\eta_{p}$ block, in the presence of an external field B, will be
\begin{equation}
 x_{\eta_{p}}=-B \sum_{i=0}^{n-1}s_{i}^{(p)}+\sum_{i=0}^{n-2}\sum_{k=i+1}^{n-1} \Lambda(s_{i}^{(p)},s_{k}^{(p)}),\label{eq:xeta}
\end{equation}
which defines a vector $\langle X|$ of length $\upsilon$. The contribution of type II will be denoted by $y_{\eta_{p}\eta_{p+1}}$, and will be given by
\begin{equation}
 y_{\eta_{p}\eta_{p+1}}=\sum_{i=0}^{n-1}\sum_{k=0}^{i} \Lambda(s_{i}^{(p)},s_{k}^{(p+1)}),\label{eq:yeta}
\end{equation}
which defines a $\upsilon\times\upsilon$ matrix. In general $Y_{\eta_{i}\eta_{j}}\neq Y_{\eta_{j}\eta_{i}}$ which makes $Y$ non symmetric.
The energy of the whole configuration $S^{L}$ can then be written as
\begin{equation}
\Lambda (s^{L})  = x_{\eta_{0}} + y_{\eta_{0}\eta_{1}} + x_{\eta_{1}} + y_{\eta_{1}\eta_{2}} + \ldots + y_{\eta_{N-2}\eta_{N-1}} + x_{\eta_{N-1}}. \label{eq:slenergy}
\end{equation}

The vector  $\langle U|$ and the matrix $V$ are then introduced as
\begin{equation}
 u_{\eta_{i}}=\exp(-\frac{1}{2}\beta x_{\eta_{i}})\label{Umatrix}
\end{equation}
\begin{equation}
v_{\eta_{i}\eta_{j}}=\exp[-\beta (\frac{1}{2} x_{\eta_{i}}+y_{\eta_{i}\eta_{j}}+\frac{1}{2}x_{\eta_{j}})].\label{Vmatrix}
\end{equation}
where $\beta\equiv(k_{B}T)^{-1}$ is the Boltzmann product. $V$ is known as the transfer matrix.

The partition function follows
\begin{equation}
\displaystyle  \begin{array}{l}Z_{Nn}=\displaystyle  \sum_{\eta_{0}=0}^{\upsilon-1} \displaystyle \sum_{\eta_{1}=0}^{\upsilon-1}\ldots \displaystyle \sum_{\eta_{N-1}=0}^{\upsilon-1}\exp[-\beta \Lambda (s^{L})] \\ \\
=\langle U | V^{N-1} | U\rangle\end{array},\label{eq:partition}
\end{equation}
for open (free) boundary conditions. For close (periodic) boundary conditions
\begin{equation}
\displaystyle  Z_{Nn}=Tr(V^{N}).\label{eq:znperiodic}
\end{equation}
$Tr(M)$ denotes the trace of the matrix $M$.

As the trace of a matrix is invariant to similarity transformations, from equation (\ref{eq:znperiodic}) for close boundary conditions
\begin{equation}
 Z_{Nn}=\sum \lambda_{i}^{N}
\end{equation}
follows. $\lambda_{i}$ are the eigenvalues of the matrix $V$. If $\lambda_{i}$ is degenerate, then the term is added as much times as its multiplicity. If the eigenvalues are labeled in non increasing order ($|\lambda_{i}| \geq |\lambda_{j}|\; \; \rightarrow j > i $), then for $N\gg 1$
\begin{equation}
 Z_{Nn}=\lambda_{0}^{N}\label{eq:ZNnciclico}
\end{equation}
where $\lambda_{0}$ is known as the dominant eigenvalue and, according to the Perron-Frobenius theorem, it is real, positive and non-degenerate\cite{behrends00}.

For open boundary conditions, using again the Perron- Frobenius theorem for a square positive defined matrix $V$, the following holds
\begin{equation}
\lim_{N \longrightarrow \infty} \frac{V^{N}}{\lambda_{0}^{N}}=|\,_{r}a_{0} \rangle \langle \,_{l}a_{0}|, \label{eq:pf}
\end{equation}
where $\langle \,_{l}a_{0} |$ y $|\,_{r}a_{0} \rangle$ are respectively, the left and right eigenvectors  corresponding to the dominant eigenvalue. The eigenvectors are normalized in the sense of $\langle \,_{l}a_{0} | \,_{r}a_{0} \rangle=1$. The matrix  $|\,_{r}a_{0} \rangle \langle \,_{l}a_{0}|$ is known as Perron projection matrix. Using (\ref{eq:pf}) and (\ref{eq:partition}) we arrive at 
\begin{equation} 
 Z_{Nn}=\langle U | \,_{r}a_{0} \rangle \langle \,_{l}a_{0}|U\rangle \lambda_{0}^{N-1}\label{eq:ZNnabierto}
\end{equation}
which in the particular case of a diagonalizable matrix reduces to
\begin{equation}
Z_{Nn}=u_{0}^{2}\langle a_{0}|a_{0}\rangle \lambda_{0}^{N-1} \label{eq:ZNnautovector}
\end{equation}
and $u_{i}$ are the components of the vector $\langle U |$ in the orthogonal base defined by the eigenvectors $\langle a_{i}|$.
It is well documented how the thermodynamic magnitudes can be obtained from the partition function \cite{lavis99}.

\section{Ising model as a Markov (Gibbs) random field}\label{sec:randomfield}

The probability of a given spin chain will be given by
\begin{equation}
\displaystyle  \begin{array}{ll}
Pr(s^{L})&=\frac{1}{Z_{Nn}}e^{-\beta \Lambda(s^{L})} \\
& \\
&=\frac{1}{Z_{Nn}}\left ( U_{\eta_{0}}V_{\eta_{0}\eta_{1}}V_{\eta_{1}\eta_{2}}\ldots V_{\eta_{N-2}\eta_{N-1}}U_{\eta_{N-1}}\right )\\
\\
&=\frac{U_{\eta_{0}} U_{\eta_{N-1}}}{M \lambda_{0}^{N-1}}\displaystyle\prod_{i=0}^{N-2}V_{\eta_{i}\eta_{i+1}}
\end{array}\label{eq:prslabierto}
\end{equation}
valid for free boundary conditions, and where we have written $M\equiv \langle U | \,_{r}a_{0} \rangle \langle \,_{l}a_{0}|U\rangle$ \footnote{Equation (7.2) in \cite{feldman98} is  similar to equation (\ref{eq:prslabierto}),  but the interpretation made of the parameters is different, as in \cite{feldman98} the product appearing in  the numerator is identified with the product of the eigenvectors, which is not the case.}. For periodic boundary conditions
\begin{equation}
\displaystyle  Pr(s^{L})=\frac{1}{\lambda_{0}^{N}}\prod_{i=0}^{N-2}V_{\eta_{i}\eta_{i+1}}\label{eq:prslciclico}
\end{equation}asked
in both cases $N\gg 1$.

Consider the spin chain divided in two half, the fist half will be called the past 
\[\displaystyle \overleftarrow{s}^{L}=s_{-L}s_{-L+1}\ldots s_{-1}=\eta_{-N}\eta_{-N+1}\ldots \eta_{-1},\]
while the second half, called the future, is given by
\[\displaystyle \overrightarrow{s}^{L}=s_{0}s_{1}\ldots s_{L-1}=\eta_{0}\eta_{1}\ldots \eta_{N-1}.\]
The spin $s_{0}$ is sometimes refered to as the present.

The known relation
\begin{equation}
\displaystyle  Pr(\overrightarrow{s}^{L}|\overleftarrow{s}^{L})=\frac{Pr(\overleftarrow{s}^{L}, \overrightarrow{s}^{L})}{Pr(\overleftarrow{s}^{L})}\label{eq:prsfuturospasado}
\end{equation}
may tempt to use equations  (\ref{eq:prslabierto}) or (\ref{eq:prslciclico}), but one must be warned against doing so. Both expressions must be understood as valid for whole systems (spin chains) and not for portions of the systems. In this way, $Pr(\overleftarrow{s}^{L})$ in the denominator of (\ref{eq:prsfuturospasado}) can not be taken as the probability of the isolated configuration $\overleftarrow{s}^{L}$ instead, use must be made of the relation
\begin{equation}
Pr(\overleftarrow{s}^{L})=\sum_{\{\overrightarrow{s}^{L}\}}Pr(\overleftarrow{s}^{L}, \overrightarrow{s}^{L}).\label{eq:marg}
\end{equation}
The reader can check that the result is not the same. This is at the heart of the failure in the mathematical treatment in [\cite{feldman98,feldman98a,crutchfield97}]  the random field character of the system is overlooked.

The Ising model is a particular case of a Gibbs random field\cite{behrends00}.  In a formal way, a set of sites $\Psi={0,1,2,\ldots , N-1}$ is given, together with a finite set $\Upsilon$ of cardinality  $\upsilon$. There is a correspondence $x:\Psi \rightarrow \Upsilon$ such that $s^L$ represents the configuration where each site $i \in \Psi$ has been assigned a value $\eta_{i}\in \Upsilon$. $\Psi^{\Upsilon}$ is the set of all possible configurations. Consider a probabilistic measure $\mathbb{P}$ associated to the space $\Psi^{\Upsilon}$. $\{\Psi^{\Upsilon}, \mathbb{P}\}$ will be called  a random field and $\eta_{i}$ will be random variables with stochastic dependence among them. For each $\eta_{i}\in \Upsilon$ the conditional probabilities  
\begin{equation}
\begin{array}{l}
Pr(\eta_{i}=\eta| \eta_{t}=b_{t}\, \text{for}\, t\in \Psi, \, t\neq i,\, b_{t}\in \Upsilon)\\\\
=Pr(\eta_{i}=\eta| s^{L}-\eta_{i}),
\end{array}\label{eq:prrandom}
\end{equation}
are well defined, where $s^{L}-\eta_{i}$ denotes the configuration $s^{L}$ excluding site $i$ ( a better notation would had been $s^{L}-i$, but the introduced one is sufficiently clear and convenient in what follows).

The case of interest is one where it is not necessary to know the whole configuration $s^{L}-\eta_{i}$ to determine the probabilities (\ref{eq:prrandom}) but it suffices to know the values $\eta_{t}$ in a subset $\mathcal{N}_{i}\subset \Psi$, which is called neighborhood and complies with the conditions
\begin{enumerate}
 \item For $i \in \Psi$, $\mathbb{N}_{i}$ is a subset (possibly empty) of $\Psi$ that does not contain $i$.
 \item $t\in \mathcal{N}_{i} \rightarrow i \in \mathcal{N}_{t}$
\end{enumerate}
A Markov random field is defined as one fulfilling
\begin{equation}
Pr(\eta_{i}=\eta| s^{L}-\eta_{i})=Pr(\eta_{i}=\eta| \eta_t =b_t\, \text{for}\, t \in \mathcal{N}_{i}\, b_{t}\in \Upsilon).\label{eq:markov}
\end{equation}
These probabilities are called the local characteristics associated to $\mathbb{P}$.

Let $\mathcal{N}_{k}={k-1,k+1}$ for $k=1,2,\ldots N-2$, $\mathcal{N}_{0}={1}$, $\mathcal{N}_{N-1}={N-2}$. Then, with respect to this system of neighborhood of the Markov random field, there is associated a Markov chain  $\{x_{i}\}_{i=0,\ldots N-1}$.  

If $s^{L}-\eta_{i}$ is the configuration $s^{L}$ without considering the block $\eta_{i}$, then, the probability $Pr(\eta_{i}|s^{L}-\eta_{i})$ that the $i$-block has value $\eta_{i}$ when all the other spins  (that is, excluding the  $\eta_{i}$ block) will have the configuration $s^{L}-\eta_{i}$ will be given by the product rule 
\begin{equation}
 Pr(\eta_{i}|s^{L}-\eta_{i})=\frac{Pr(s^{L})}{Pr(s^{L}-\eta_{i})}=\frac{Pr(s^{L})}{\sum_{s^{L*}}Pr(s^{L*})}\label{eq:prcaterva}
\end{equation}
where the sum $s^{L*}$ is over all configurations identical to $s^{L}$ except, possibly, for the block $\eta_{i}$.

Using equation (\ref{eq:prslabierto}), the probability of a configuration will be  
\begin{equation}
\displaystyle\begin{array}{ll}
 Pr(s^{L})=&\frac{1}{Z_{Nn}}e^{-\beta \Lambda(s^{L})}\\\\
&=\frac{1}{Z_{Nn}}\exp \left [-\beta \left ( \sum_{i=0}^{N-1}x_{\eta_{i}}+\sum_{i=0}^{N-2}y_{\eta_{i}\eta_{i+1}}\right ) \right ]\\\\
&\displaystyle=\frac{1}{Z_{Nn}} e^{-\beta x_{\eta_{N-1}}}\prod_{j=1}^{N-2}e^{-\beta x_{\eta_{j}}} e^{-\beta y_{\eta_{j}\eta_{j+1}}}
\end{array}
\end{equation}
and,
\begin {equation}
\begin{array}{l}
 \displaystyle Pr(s^{L}-\eta_{i})=\sum_{s^{L*}}Pr(s^{L*})=\\\\
\displaystyle =\left [ \frac{1}{Z_{Nn}} e^{-\beta x_{\eta_{N-1}}}\prod_{\overset{j=1}{j\neq i; \, j\neq i-1}}^{N-2}e^{-\beta x_{\eta_{j}}} e^{-\beta y_{\eta_{j}\eta_{j+1}}} \right ] e^{-\beta x_{\eta_{i-1}}}\\\\
\displaystyle \sum_{\eta_{k}}e^{-\beta y_{\eta_{i-1}\eta_{k}}} e^{-\beta x_{\eta_{k}}} e^{-\beta y_{\eta_{k}\eta_{i+1}}}.
\end{array}
\end {equation}
For the local characteristics equation (\ref{eq:prcaterva}) will now be 
\begin{equation}
\begin{array}{ll}
 Pr(\eta_{i}|s^{L}-\eta_{i})&=\frac{e^{-\beta y_{\eta_{i-1}\eta_{i}}} e^{-\beta x_{\eta_{i}}} e^{-\beta y_{\eta_{i}\eta_{i+1}}}}{\sum_{\eta_{k}}e^{-\beta y_{\eta_{i-1}\eta_{k}}} e^{-\beta x_{\eta_{k}}} e^{-\beta y_{\eta_{k}\eta_{i+1}}}}\\\\
&=\frac{V_{\eta_{i-1}\eta_{i}}V_{\eta_{i}\eta_{i+1}}}{\sum_{\eta_{k}} V_{\eta_{i-1}\eta_{k}}V_{\eta_{k}\eta_{i+1}} }
\end{array}\label{eq:caractloc}
\end{equation}
for blocks $\eta_{i}$ not at the extremes. 

For the first block
\begin{equation}
\begin{array}{ll}
 Pr_{0}(\eta_{0}|s^{L}-\eta_{i})&=\frac{e^{-\beta x_{\eta_{0}}} e^{-\beta y_{\eta_{0}\eta_{1}}}}{\sum_{\eta_{k}} e^{-\beta x_{\eta_{k}}} e^{-\beta y_{\eta_{k}\eta_{1}}}}\\\\
&=\frac{U_{\eta_{0}}V_{\eta_{0}\eta_{1}}}{\sum_{\eta_{k}} U_{\eta_{k}}V_{\eta_{k}\eta_{1}}}
\end{array}\label{eq:caractloceta0}
\end{equation}
Similar expression can be found for the last block.
Expression (\ref{eq:caractloc}) has the important consequence that
\begin{equation}
 Pr(\eta_{i}|s^{L}-\eta_{i})=Pr(\eta_{i}|\eta_{i-1},\eta_{i+1})
\end{equation}

It must be noted that one of the consequence of equations (\ref{eq:caractloc}) and (\ref{eq:caractloceta0}), is that the probability of $\eta_{i}$ taking a particular value is conditional on the values of the immediate neighboring blocks, and not by the whole configuration $s^{L}-\eta_{i}$, this determines the Markov character of the random field. 

The stochastic matrix $P$ will be defined by
\begin{equation}
 P_{ij}=Pr(\eta_{j}|\eta_{i}).
\end{equation}
which defines a transition probability from state $\eta_i$ to state $\eta_j$. By definition $\sum_{j}P_{ij}=1$. If $\langle p^{\infty}|$ is the vector of probabilities over the blocks $\eta_{i}$ then is well known that the stationary distribution\cite{behrends00} is given by 
\begin{equation}
\langle p^{\infty}|=\langle w_{0}| \label{eq:prstac}
\end{equation}
where $\langle w_{0}| $ is the left dominant eigenvector of the matrix $P$.
The vector $\langle p^{\infty}|$ allows to calculate $Pr(\eta_{i})$ when the Markov process has been running for a sufficiently long time. 

The local characteristics can be written in terms of the stochastic matrix $P$ using Bayes theorem 
\begin{equation}
 \begin{array}{l}
  Pr(\eta_{i}, \eta_{i-1},\eta_{i+1})=Pr(\eta_{i}| \eta_{i-1},\eta_{i+1})Pr(\eta_{i+1},\eta_{i-1})=\\\\
=Pr(\eta_{i}| \eta_{i-1},\eta_{i+1})Pr(\eta_{i+1}|\eta_{i-1})Pr(\eta_{i-1})
 \end{array}\label{eq:protra}
\end{equation}
similarly
\begin{equation}
 \begin{array}{l}
  Pr(\eta_{i}, \eta_{i-1},\eta_{i+1})=Pr(\eta_{i+1}| \eta_{i},\eta_{i-1})Pr(\eta_{i},\eta_{i-1})=\\\\
=Pr(\eta_{i+1}| \eta_{i},\eta_{i-1})Pr(\eta_{i}|\eta_{i-1})Pr(\eta_{i-1})
 \end{array}\label{eq:protra1}
\end{equation}
now equating both (\ref{eq:protra}) and (\ref{eq:protra1})
\begin{equation}
 \begin{array}{ll}
  Pr(\eta_{i}|\eta_{i-1},\eta_{i+1})&=\frac{Pr(\eta_{i}|\eta_{i-1})Pr(\eta_{i+1}|\eta_{i}\eta_{i-1})}{Pr(\eta_{i+1}|\eta_{i-1})}\\\\
&=\frac{Pr(\eta_{i}|\eta_{i-1})Pr(\eta_{i+1}|\eta_{i})}{\sum_{l} Pr(\eta_{l}|\eta_{i-1})Pr(\eta_{i+1}|\eta_{l})}
 \end{array}\label{eq:localcstochstic}
\end{equation}
where the Markov character of the field has been used and the last step is justified by the total probability theorem\footnote{This last equation together with eq. (\ref{eq:caractloceta0}) shows that equation (7.6) and (7.15) in \cite{feldman98} are flawed.}.

Equation (\ref{eq:localcstochstic}) can be rewritten as
\begin{equation}
  Pr(\eta_{i}|\eta_{i-1},\eta_{i+1}) \sum_{l} Pr(\eta_{l}|\eta_{i-1})Pr(\eta_{i+1}|\eta_{l})=Pr(\eta_{i}|\eta_{i-1})Pr(\eta_{i+1}|\eta_{i})\label{eq:localcstochstic1}
\end{equation}
which forms, when written for each $\eta_{i}$, an homogeneous system of quadratic forms. Such system can have non trivial solution if it is undetermined, which happens in this case if  the square of the number of unknown is larger than the number of equations.

As each local characteristic is determined by three $\eta$'s, there will be $\theta^{3n}=\nu^{3}$ equations and $\theta^{2n}=\nu^{2}$ unknowns. The relations 
\[\displaystyle \sum_{i}^{\nu} Pr(\eta_{i}| \eta_{j})=1\;\;\;\;\;\;\forall j \]
must be added that eliminates $\nu$ unknowns. The total number of unknowns is then $\nu(\nu-1)$ and the total number of equations $\nu^{3}$.
\[\nu^{2}(\nu-1)<\nu^{2}(\nu-1)^{2}\Longrightarrow \nu > 2.\]
For $\nu=2$, the system will also have non trivial solution as a result of the additional reduction of equations by the symmetry of the transfer matrix. Returning to equation (\ref{eq:localcstochstic1}) and rewriting for any local characteristic
\begin{equation}
  Pr(\eta_{i}|\eta_{j},\eta_{m}) \sum_{l} Pr(\eta_{l}|\eta_{j})Pr(\eta_{m}|\eta_{l})=Pr(\eta_{i}|\eta_{j})Pr(\eta_{m}|\eta_{i}),\label{eq:localcstochstic2}
\end{equation}
introducing
\begin{equation}
 Y_{l}(j,m)=Pr(\eta_{l}|\eta_{j})Pr(\eta_{m}|\eta_{l})\label{eq:yfunc}
\end{equation}
equation (\ref{eq:localcstochstic2}) can be written as
\begin{equation}
  Pr(\eta_{i}|\eta_{j},\eta_{m}) \sum_{l} Y_{l}(j,m)=Y_{i}(j,m).\label{eq:localcstochstic3}
\end{equation}
The normalization condition ( which can be derived from equation (\ref{eq:localcstochstic})) over the local characteristics determines 
\begin{equation}
 \displaystyle \sum_{k}^{\nu}Pr(\eta_{k}|\eta_{j},\eta_{m})=1,\label{eq:prnorm}
\end{equation}
Equation (\ref{eq:localcstochstic3}) is linear and homogeneous over the $Y_{i}(j,m)$ which, upon solving for the non trivial-case, leads to a system of simple homogeneous quadratic equations (\ref{eq:yfunc}) which can be readily solved.

The Markov character of the system means that the generation process can forget of all the past except the last block  $\eta_{-1}$ (the last $n$ spins), to determine, as certain as possible the future. In other words, if the local characteristics implies a stochastic matrix as equation (\ref{eq:localcstochstic2}) imply, then all past configuration $\overleftarrow{s}^{L}$ with the same last block $\eta_{-1}$ conditions (statistically) the same future, this fact allows to consider the Ising chain as a canonical finite state machine known as $\epsilon$-Machine.

\section{Ising model as a canonical finite state machine: $\epsilon$-Machine}\label{sec:epsilonmachine}

Causal states are at the core of the computational mechanic framework\cite{shalizi01}. If two blocks $\eta_{-1}$ and $\eta'_{-1}$ give the same $Pr(\overrightarrow{s}^{L}|\overleftarrow{s}^{L})$, for all possible futures $\overrightarrow{s}^{L}$, then $\eta_{-1}$ y $\eta'_{-1}$ are said to belong to the same causal state ($C_{p}$) and we write $\eta_{-1}\sim \eta'_{-1}, \eta_{-1}, \, \eta'_{-1}\in C_{p} $. Two blocks belonging to the same causal state $C_{p}$ define identical rows in the stochastic matrix.

The partition of the set $\Upsilon$ in classes of causal states is an equivalence relation complying with the transitivity condition (if $\eta_{i} \sim \eta_{j}$ y $\eta_{j} \sim \eta_{k}$ then $\eta_{i} \sim \eta_{k}$), symmetry (if $\eta_{i} \sim \eta_{j}$  then  $\eta_{j} \sim \eta_{i}$ )  and reflectivity ($\eta_{i} \sim \eta_{i}$).  The set of causal state uniquely determines the future of a sequence. The set of all causal states will be denoted by $\mathcal{C}$ with cardinality $|\mathcal{C}|$. A function $\epsilon$ can be defined over $\eta$, $\epsilon: \Upsilon \rightarrow \mathcal{C}$ which relates $\eta$ with its causal states $C$, 
\[
\epsilon(\eta)\equiv C=\{\eta'| Pr(\overrightarrow{s}|\eta)=Pr(\overrightarrow{s}|\eta')\;\; \forall \overrightarrow{s}\}.
\]

The probability of a (recursive) causal state  is directly deducible from equation (\ref{eq:prstac}),
\begin{equation}
 \displaystyle Pr(C_{p})=\sum_{\eta_{j}\in C_{p}} Pr(\eta_{j})=\sum_{\eta_{j}\in C_{p}} p^{\infty}_{\eta_{j}}\label{eq:prcp}
\end{equation}
As each causal state represents the set of past that determines (probabilistically) the same future, the set of causal state represents the memory the system has to keep in order to predict, as good as possible, the future.

The statistical complexity has been defined as the Shannon entropy over the causal states\cite{crutchfield94}
\begin{equation}\begin{array}{ll}
 \displaystyle C_{\mu}&\equiv - \sum_{C_{p}\in\mathcal{C}}Pr(C_{p})\log Pr(C_{p})\\\\
&=H[\mathcal{C}].
\end{array}\label{eq:Cmu}
\end{equation}
The logarithm is usually taken in base two and the units are then bits. Statistical complexity is a measure of how much memory (resources) the system needs to optimally predict the future.  If the system has $|C|$ causal states then the statistical complexity has the upper bound
\[C_{\mu} \leq \log |C|,\]
corresponding to a uniform distribution of probabilities over the causal states. The relation follows from a known property of the Shannon entropy. The upper bound of the statistical complexity is also known as topological entropy. As already stated, for a fixed number of causal states, the increase of statistical complexity is the result of a more uniform distribution over the causal state, in this sense such increase witness the increase of uncertainty over the occurrence of such states and therefore the unreductible entropy of the system (e.g.  thermal noise). The system needs more resources to account for the unpredictable behavior of the system.

The probability of occurrence of block $\eta_{i}$ conditional on the causal state $C$ will be given by
\begin{equation}
\begin{array}{ll}
 Pr(\eta_{i}|C)&=\sum_{\eta_{k}} Pr(\eta_{i}|\eta_{k} \cap C)Pr(\eta_{k}| C)\\\\
&=Pr(\eta_{i}|\eta_{j}; \eta_{j}\in C) \sum_{\eta_{k}\in C}Pr(\eta_{k}|C)\\\\
&=Pr(\eta_{i}|\eta_{j}; \eta_{j}\in C)
\end{array}\label{eq:pretac}
\end{equation}
in the first step the total probability theorem was used, in the second step use has been made of the fact that conditioning in $\eta_{k}\cap C$ is equal to conditioning in $\eta_{k}$ if the block belongs to the causal state $C$ and, finally $\sum_{\eta_{k}\in C}Pr(\eta_{k}|C)=1$.
\[
 \begin{array}{ll}
  Pr(\eta_{j})&=\sum_{\eta_{k}\in\Sigma}Pr(\eta_{j}|\eta_{k})Pr(\eta_{k})\\\\
&=\sum_{C_{k}\in\mathcal{C}}\sum_{\eta_{k}\in C_{k}}Pr(\eta_{j}|\eta_{k})Pr(\eta_{k})\\\\
&=\sum_{C_{k}\in\mathcal{C}}Pr(\eta_{j}|\eta_{k'}, \eta_{k'} \in C_{k}) \sum_{\eta_{k}\in C_{k}}Pr(\eta_{k})\\\\
&=\sum_{C_{k}\in\mathcal{C}}Pr(\eta_{j}|\eta_{k'}, \eta_{k'} \in C_{k}) Pr(C_{k})\\\\
 \end{array}
\]
now we make use of equation (\ref{eq:pretac}) to get
\begin{equation}
 Pr(\eta_{j})=\sum_{C_{k}\in\mathcal{C}} Pr(\eta_{j}| C_{k})Pr(C_{k}).
\end{equation}
which allows to compute the occurrence of a block from the probabilities over the causal states.

The probability of a transition from one causal state to other will be given by
\begin{equation}
Pr(C_{k}\rightarrow C_{p})\equiv Pr(C_{p}|C_{k})=\sum_{\eta_{j}\in C_{p}} Pr(\eta_{j}| C_{k})
\end{equation}

If we define the transition matrix $T^{(\eta)}$, whose elements are the probability of going from state $C_{k}$ to state $C_{p}$ upon emitting a block $\eta$:
\begin{equation}
 T_{rq}^{(\eta)}=Pr(C_{r}\overset{\eta}{\rightarrow}C_{q}).
\end{equation}
By construction, the emission of a block $\eta$ determines uniquely the causal state to where the transition occurs (this is called the unifiliar property\cite{shalizi01}). In this sense, the generation process is deterministic.
Correspondingly, the connectivity matrix $T$  is defined as
\begin{equation}
 \displaystyle T_{rq}=\sum_{\eta\in \Sigma}T_{rq}^{(\eta)}=Pr(C_{r}\rightarrow C_{q})
\end{equation}
which connects causal states without regard of the emitted block.

An $\epsilon$-machine is defined by the causal state function $\epsilon$, which relates histories with causal states and the transition matrices between them. In the case of the Ising model, the $\epsilon$-machine is represented by a deterministic finite state machine with vertex defined by the causal states and edges labeled according to the transition probabilities\cite{crutchfield97}. 

In order to account for the irreducible randomness, the density of entropy is defined as\cite{feldman08}
\begin{equation}
\begin{array}{ll}
 \displaystyle h_{\mu}&=\lim_{L\rightarrow \infty} H[\eta_{0}|\eta_{-L}\eta_{-L+1}\ldots \eta_{-1}]\\\\
&=\lim_{L\rightarrow \infty} H[\eta_{0}|\overleftarrow{s}^{L}]
\end{array}\label{eq:hmiubloque}
\end{equation}
$h_{\mu}$  is the uncertainty on the next emitted block $\eta_{0}$ conditional on having seen infinite previous blocks (spins). 
By definition $h_{\mu} \geq 0$.
\begin{equation}
\displaystyle
\begin{array}{ll}
 H[\eta_{0}|\overleftarrow{s}^{L}]&=-\sum_{\eta_{0}}\sum_{\eta_{-n}}\ldots \sum_{\eta_{-1}} Pr(\overleftarrow{s}^{L}\eta_{0})\log Pr(\eta_{0}|\overleftarrow{s}^{L})\\
\\
&=-\sum_{\eta_{0}}\sum_{\eta_{-N}}\ldots \sum_{\eta_{-1}} Pr(\eta_{0}|\overleftarrow{s}^{L}) Pr(\overleftarrow{s}^{L})\log Pr(\eta_{0}|\overleftarrow{s}^{L})
\end{array}\label{eq:entropyrate1}
\end{equation}
where $\overleftarrow{s}^{L}=\eta_{-N}\eta_{-N+1}\ldots \eta_{-1}$, , on the other hand, using $\overleftarrow{s}^{L-n}=\eta_{-N}\ldots \eta_{-2}$ 
\begin{equation}
\displaystyle \begin{array}{ll}
 Pr(s^{L})&=Pr(\eta_ {-1}|s^{L-n})P(s^{L-n})\\
\\
&=Pr(s^{L-n}|\eta_{-1})Pr(\eta_{-1}).\label{eq:prslbayesbloque}
\end{array}
\end{equation}
In the last step, use has been made of Bayes theorem. Substituting equation (\ref{eq:prslbayesbloque}) on equation (\ref{eq:entropyrate1}) and reordering terms
\begin{equation}
\displaystyle \begin{array}{l}
 H[\eta_{0}|\overleftarrow{s}^{L}]=-\sum_{\eta_{0}}\left [ \sum_{\eta_{-1}} Pr(\eta_{0}|\eta_{-1})Pr(\eta_{-1})\log Pr(\eta_{0}|\eta_{-1}) \right.\\
\\ \left. \left \{ \sum_{\eta_{-N}}\ldots \sum_{\eta_{-2}}Pr(\overleftarrow{s}^{L-n}|\eta_{-1})\right \}\right ],
\end{array}\label{eq:Hsslbloque}
\end{equation}
now
\[\displaystyle  \sum_{\eta_{-N}}\ldots \sum_{\eta_{-2}}Pr(\overleftarrow{s}^{L-n}|\eta_{-1}) \]
is the probability that from $\eta_{-1}$ any configuration is conditioned and that probability is $1$. Equation (\ref{eq:Hsslbloque}) then reduces to
\begin{equation}
\begin{array}{ll}
 \displaystyle  h_{\mu}&=\lim_{L\rightarrow\infty}H[\eta_{0}|\overleftarrow{s}^{L}]=H[\eta_{0}|\eta_{-1}]\\\\
&=-\sum_{\eta_{j}\in \Sigma}Pr(\eta_{j})\sum_{\eta_{i}\in \Sigma}Pr(\eta_{i}|\eta_{j})\log Pr(\eta_{i}|\eta_{j})\\\\
&=-\sum_{C_{\alpha}\in \mathcal{C}} Pr(C_{\alpha})\sum_{\eta_{k} \in \Sigma}Pr(\eta_{k}|C_{\alpha})\log Pr(\eta_{k}|C_{\alpha})
\end{array}\label{eq:entropyrateblock}
\end{equation}
 
The mutual information between past and future is called the excess entropy\cite{feldman08}
\begin{equation}
 E_{\mu}\equiv I [\overleftarrow{s}:\overrightarrow{s}],
\end{equation}
where $I[X:Y]$ is the mutual information between $X$ e $Y$. From the finite range character of the interaction in the Ising model
\begin{equation}
\displaystyle 
\begin{array}{ll}
E_{\mu}= I [\overleftarrow{s}:\overrightarrow{s}]&=I[\eta_{-1}:\eta_{0}]\\
&\\
&=H[\eta_{-1}]-H[\eta_{0}|\eta_{-1}]
\end{array}\label{eq:excess}
\end{equation}
where
\begin{equation}
\displaystyle H[\eta_{i}]=\sum_{\eta_{i}\in \Sigma}Pr(\eta_{i})\log Pr(\eta_{i}),\label{eq:Hetai}
\end{equation}
and
\[
H[\eta_{0}|\eta_{-1}]=H[\eta_{0}|\overleftarrow{s}^{L}]=h_{\mu}
\]
given by equation  (\ref{eq:entropyrateblock}).

From equation (\ref{eq:Cmu}) and (\ref{eq:entropyrateblock}) we arrive to the expression
\begin{equation}
 E_{\mu}=C_{\mu}-h_{\mu}.
\end{equation}
Excess entropy is a measure of the resources needed by the system in order to optimally predict the future, once the irreducible randomness has been subtracted\cite{feldman08}. As $E$ is a mutual information, it will always be a non-negative value, which implies
\[ C_{\mu}\geq h_{\mu}\]
If the system is completely ordered then $h_{\mu}=0$ and
\[ C_{\mu}=E_{\mu}.\]

\subsection{Markov process as emission of single spins}

The result, that in order to correctly derive the stochastic matrix, one needs the local characteristic of the Markov field, points to fact that is not straight forward to cast a spacially extended process into a sequencial process. In the Markov field, all spin values over the whole one dimensional lattice is given simultaneously. In general, the spin value at one site depends on the spin values to the right, as well as to the left, within the interaction range.  It is the transfer matrix method that rewrites the whole system as a one over blocks of size equal to the interaction range, that allows to cast the dynamics of the system as Markov process, but in doing so, the building blocks cease to be the individual spins and instead the $\eta$ blocks are the new individual entities. It is then, not straight forward to consider valid, if one can describe the $\epsilon$-machine process over single spin emission. This is contrast with previous treatment of the subject \cite{crutchfield97,feldman98} and some general conclusion derived from it \cite{feldman03}. To exemplify the problems  associated with defining the Markov process over the emission of a single spin, consider a general Hamiltonian with interaction range $n=2$, in such system a eight period perfect sequence is possible with probability larger than zero. The four state FSM with single spin emission is incapable of topologically reproducing such sequence. There is then the need to introduce ``bogus'' states additional to the $\eta$ blocks.  Yet, the stochastic matrix is well defined by the local characteristic over the $\eta$ blocks and the $\epsilon$-machine over block emission makes the introduction of unnatural states unnecessary.

Having said so, in what follows the Markov process will be formally cast in terms of a single spin process to bridge the gap with the mathematical developments made before. In the previous treatment, the Markov process has been constructed taking as a single event the emission of a whole block $\eta$ of length $n$. Other treatments have considered the Markov process single event the emission of a single spin\cite{feldman98,feldman98a,crutchfield97}.
Let us introduce the operator $\pi$ such that if $\eta_{k}=s_{k}s_{k+1}\ldots s_{k+n-2}s_{k+n-1}$, then
\[
\pi\eta_{k}=s_{k+1}\ldots s_{k+n-1}
\]
and $\pi\eta_{k}$ has length $n-1$. Similarly
\[
\eta_{k}\pi=s_{k}\ldots s_{k+n-2}
\]

The probability of emitting a single spin $s_{0}=s$ conditioned in being on the causal state $C$ will be denoted by $Pr(s| C)$. Using equation (\ref{eq:pretac}),
\begin{equation}
\begin{array}{cc}
 \displaystyle Pr(s|C)&=\sum_{s_{1}}\sum_{s_{2}}\ldots \sum_{s_{n-1}} Pr(\eta_{k}|\eta_{j}; \, \eta_{j}\in C)\\\\
&=\sum_{\overset{\eta_{k}\in \Upsilon}{s_{0}=s}} Pr(\eta_{k}|\eta_{j}; \, \eta_{j}\in C)
\end{array}\label{eq:prsc}
\end{equation}
and the sum is over all configurations $\eta_{i}$ with spin $s_{0}=s$. It is important to notice that upon emitting a spin $s$ a transition occurs from $\overleftarrow{s}^{L}$ to $\overleftarrow{s}^{L}s$ which implies a new causal state to which
\[
\eta_{0}=\pi \eta_{-1} s=s_{-n+1}\ldots s_{-1}(s_{0}=s)
\]
belongs and which is uniquely determined by the generated spin. The transition $\eta_{j}\rightarrow \pi \eta_{j} s$ conditional on the emission of spin $s$, $Pr(\pi \eta_{j} s|s,\eta_{j})$, is equal to 1 if the transition is allowed. 

If $\eta_{i}$ and $\eta_{j}$ are taken to belong to the same causal state $C$, then by construction
\begin{equation}
 Pr( \overrightarrow{S} \in sF|\eta_{i})=Pr(\overrightarrow{S} \in sF|\eta_{j}),
\end{equation}
where $F$ is the set of all possible futures and $sF$ is the operation of prefixing to each future the spin $s$. It is clear that $\overrightarrow{S}\in sF$ can be split into events: $s_{0}=s$ and $\pi \overrightarrow{S}\in F$, then
\begin{equation}
\begin{array}{l}
 Pr( \overrightarrow{S}\in sF|\eta_{i})=Pr(\overrightarrow{S}\in sF|\eta_{j})\\\\
Pr( s_{0}=s, \pi \overrightarrow{S}\in F|\eta_{i})=Pr(s_{0}=s, \pi\overrightarrow{S} \in F|\eta_{j}).
\end{array}
\end{equation}
Using the product rule on both sides and taking into account that for three random variables  $X$, $Y$ y $Z$,  $Pr(Z \in A, Y=y | X=x)=Pr(Z \in A| Y=y, X=x)Pr(Y=y|X=x)$,
\begin{equation}
\begin{array}{l}
Pr( \pi \overrightarrow{S}\in F|s_{0}=s, \eta_{i}) Pr(s_{0}=s| \eta_{i})=\\\\
=Pr(\pi\overrightarrow{S} \in F|s_{0}=s, \eta_{j})Pr(s_{0}=s| \eta_{j}),
\end{array}
\end{equation}

and because $Pr(s_{0}=s| \eta_{i})=Pr(s_{0}=s| \eta_{j})$ is defined strictly positive,
\begin{equation}
\begin{array}{l}
Pr( \pi \overrightarrow{S}\in F|s_{0}=s, \eta_{i})=Pr(\pi\overrightarrow{S} \in F|s_{0}=s, \eta_{j})\\\\
Pr( \pi \overrightarrow{S}\in F| \pi\eta_{i}s)=Pr(\pi\overrightarrow{S} \in F|\pi\eta_{j}s)
\end{array}\label{eq:equivs}
\end{equation}
Expression (\ref{eq:equivs}) is equivalent to say that $\pi\eta_{i}s$ and $\pi\eta_{j}s $ belong to the same causal state and therefore, the emission of one spin determines uniquely the causal state transition.

Therefore, the following holds
\[Pr(C_{k}\overset{s}{\rightarrow}C_{i})=Pr(s|C_{k}).\] 
Now, the transition matrix $T^{(s)}$ whose elements are the probability of making a transition from state $C_{k}$ to state $C_{p}$ upon emitting a spin  $s$, will be:
\begin{equation}
 T_{rq}^{(s)}=Pr(C_{r}\overset{s}{\rightarrow}C_{q})=Pr(s|C_{r}),
\end{equation}
and $\pi\eta_{r}s\in C_{q}$, $\eta_{r}\in C_{r}$.
The transition matrix between causal states regardless of the emitted symbol will be given by
\begin{equation}
 \displaystyle T_{rq}^{(s)}=\sum_{\overset{\pi\eta_{r}s\in C_{q}}{\eta_{r}\in C_{r}}}Pr(s|C_{r}),
\end{equation}
where the left dominant eigenvector determines the probability of the causal state (equation (\ref{eq:prstac}) and (\ref{eq:prcp})). The statistical complexity $C'_{\mu}$ follows.

The entropy density will now be 
\begin{equation}
 \displaystyle h'_{\mu}=\lim_{L\rightarrow \infty} H[s|s_{-L}s_{-L+1}\ldots s_{-1}]
\end{equation}
$h'_{\mu}$  is now the uncertainty in the next emitted spin $s$ conditional of having observed infinite preceding spins in the past. Now
\begin{equation}
\begin{array}{ll}
 \displaystyle  h'_{\mu}&=\lim_{L\rightarrow\infty}H[s|\overleftarrow{s}^{L}]=H[s|\eta_{-1}]\\\\
&=-\sum_{C_{\alpha}\in \mathcal{C}} Pr(C_{\alpha})\sum_{s \in \Theta}Pr(s|C_{\alpha})\log Pr(s|C_{\alpha})
\end{array}
\end{equation}

The excess entropy will now be
\begin{equation}
 E'_{\mu}=C'_{\mu}-n h'_{\mu},
\end{equation}
$n$ is the interaction range (the size of the $\eta$  blocks).

\section{Examples}\label{sec:example}
\subsection{1/2 nearest neighbors spin chain}

The 1/2-spin Ising model for the nearest neighbor interaction is defined by the interaction Hamiltonian\cite{morita72}
\begin{equation}
 E=-B \sum_{i}s_{i}-J\sum_{j}s_{j}s_{j+1},\label{eq:nnhamiltonian}
\end{equation}
where $B$ is the external field, and $J$ is the interaction parameter. Both parameters are independent. The $\eta$ blocks set will be
\[\eta={\downarrow,\uparrow}.\]

The local characteristics derived from equation (\ref{eq:caractloc}) reduce to
\begin{equation}
 \begin{array}{l}
  Pr(\downarrow|\downarrow,\downarrow)=\frac{e^{4 \beta  J}}{e^{2 \beta  B}+e^{4 \beta  J}}\\\\
  Pr(\downarrow|\downarrow,\uparrow)=\frac{1}{e^{2 \beta  B}+1}\\\\
  Pr(\uparrow|\downarrow,\downarrow)=\frac{1}{e^{4 \beta  J-2 \beta  B}+1}\\\\
  Pr(\uparrow|\downarrow,\uparrow)=\frac{e^{2 \beta  B}}{e^{2\beta  B}+1}\\\\
  Pr(\downarrow|\uparrow,\downarrow)=\frac{1}{e^{2 \beta  B}+1}\\\\
  Pr(\downarrow|\uparrow,\uparrow)=\frac{1}{e^{2 \beta  (B+2 J)}+1}\\\\
  Pr(\uparrow|\uparrow,\downarrow)=\frac{e^{2 \beta  B}}{e^{2 \beta  B}+1}\\\\
  Pr(\uparrow|\uparrow,\uparrow)=\frac{e^{2 \beta  (B+2 J)}}{e^{2\beta  (B+2 J)}+1}.
 \end{array}
\end{equation}
Solving the linear system of equation (\ref{eq:localcstochstic3}), results in the system of quadratic equations
\begin{equation}
 \begin{array}{l}
  Pr(\downarrow|\uparrow) Pr(\uparrow|\downarrow)= e^{2 \beta  B-4 \beta  J} Pr(\downarrow|\downarrow)\\\\
  Pr(\uparrow|\uparrow) = e^{2 B \beta} P(\downarrow|\downarrow),
 \end{array}
\end{equation}
together with the normalization conditions
\begin{equation}
 \begin{array}{l}
  Pr(\downarrow|\uparrow) + Pr(\uparrow|\uparrow)=1\\\\
  Pr(\downarrow|\downarrow) + Pr(\uparrow|\downarrow)=1
 \end{array}
\end{equation}
lead to the solution for\footnote{This result is equivalent to the one reported as equation (7.31) in \cite{feldman98}, but, if one calculates all matrix entry from equation (7.15), the row normalization condition is violated. Therefore, the correctness of the entries in (7.31) in \cite{feldman98} is accidental as result of forcing row normalization.}
\begin{equation}
 Pr(\uparrow|\uparrow)=\frac{2 e^{2 \beta  J}}{\sqrt{4 e^{2 \beta  B}+e^{4 \beta  (B+J)}-2 e^{2
   \beta  (B+2 J)}+e^{4 \beta  J}}+e^{2 \beta  (B+J)}+e^{2 \beta  J}}.
\end{equation}
And the other  entries of the stochastic matrix follows.

\begin{figure}[!t]
\centering
\includegraphics[scale=1]{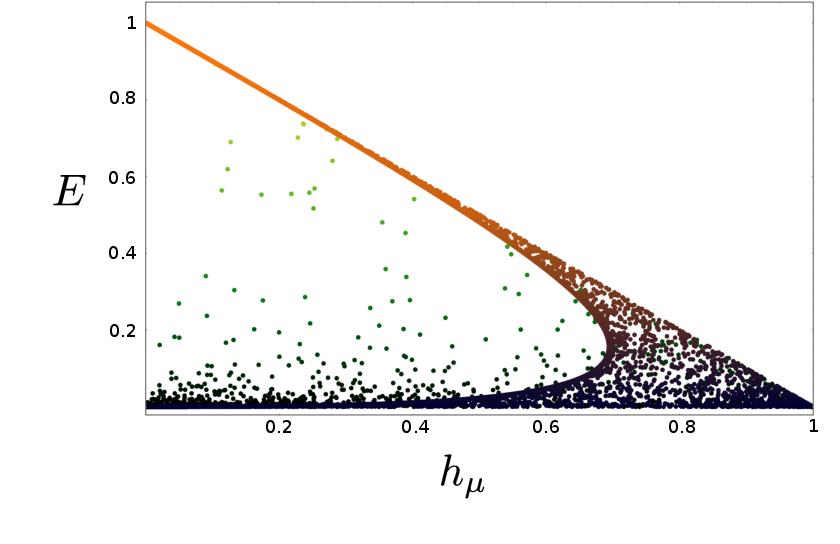}
\caption{The complexity-entropy diagram for the $1/2$-nearest neighbor Ising model. $10^5$ points were used with random parameters in the range $\beta \in [10^{-4},10^2]$, $J \in [-1.5, 1.5]$ and $B \in [-3,3]$. 
}\label{fig:nnising}
\end{figure}

From the stochastic matrix, the two state $\epsilon$-machine was built and $(h_{\mu}, E)$ were calculated for $10^5$ points taking randomly the value of the parameters. The corresponding plot is shown in figure \ref{fig:nnising}. This type of complexity map has been discussed before\cite{feldman08}. Small values of disorder can accommodate a large range of excess entropy values, which means varying probability between the two possible causal states. As disorder increases, the system looses structure tending towards a single state process which although increasingly random is also increasingly less complex.

\subsection{1/2 next nearest neighbors spin chain}

The 1/2-spin Ising model for the next nearest neighbor interaction is defined by the interaction Hamiltonian\cite{morita72}
\begin{equation}
 E=-B \sum_{i}s_{i}-J_{1}\sum_{j}s_{j}s_{j+1}-J_{2}\sum_{k}s_{k}s_{k+2}.\label{eq:nnnhamiltoniom the an}
\end{equation}
where $B$ is the external field, and $J_{1}$, $J_{2}$ are spin coupling parameters. The $\eta$ blocks set will be
\[\eta={\downarrow\downarrow, \downarrow\uparrow,\uparrow\downarrow,\uparrow \uparrow}\]
we can take the value $1$ to correspond to spin up, whereas $-1$ corresponds to spin down.

\begin{figure}[!t]
\centering
\includegraphics*[scale=1]{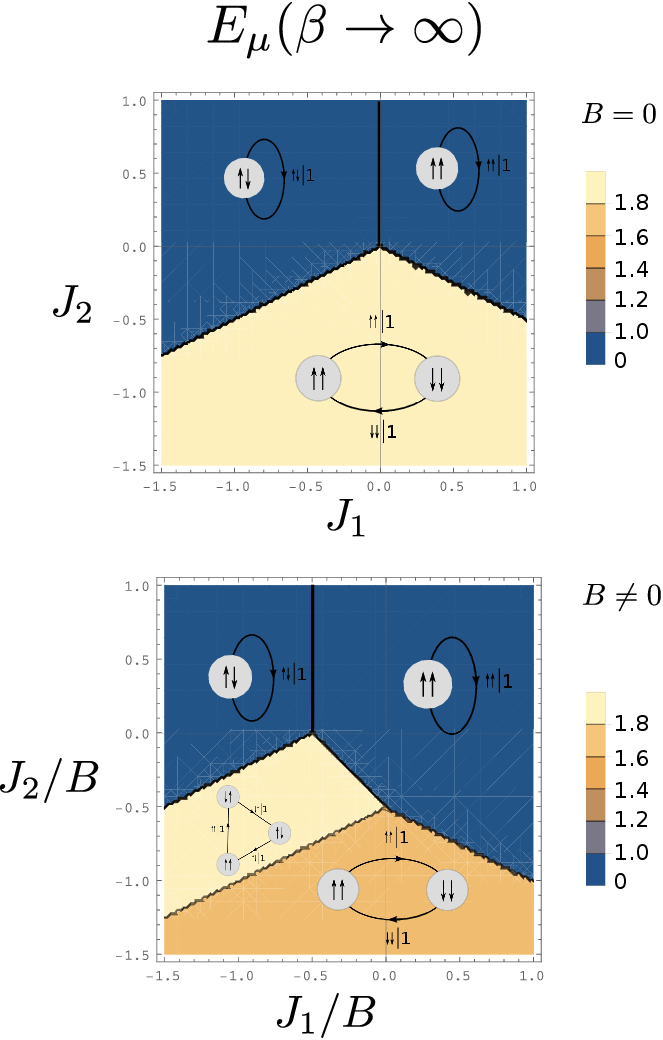}
\caption{Excess entropy ($E_{\mu}$) at the ground state ($\beta \rightarrow \infty$) for zero field ($B=0$) and non-zero field ($B\neq 0$).
}\label{fig:nnn}
\end{figure}

Figure \ref{fig:nnn} shows the excess entropy as a function of $J_{1}$ and $J_{2}$ for the ground state ($\beta \rightarrow \infty$) at zero field ($B=0$) and as a function of $J_{1}/B$ and $J_{2}/B$ for non-zero field ($B\neq 0$). The excess entropy together with the $\epsilon$-machine reconstruction allows to distinguish four orderings of periodicity 1,2,3 and 4, identified as the ordered sequences 
\[
\begin{array}{ll}
(1) & \uparrow\uparrow\uparrow\uparrow\uparrow\uparrow\uparrow\uparrow\cdots\\\\
(2) & \uparrow\downarrow\uparrow\downarrow\uparrow\downarrow\uparrow\downarrow\cdots \\\\
(3) & \uparrow\uparrow\downarrow\uparrow\uparrow\downarrow\uparrow\uparrow\cdots\\\\
(4) & \uparrow\uparrow\downarrow\downarrow\uparrow\uparrow\downarrow\downarrow\cdots
\end{array}
\]
which are the expected phases\cite{morita72}. The pure ferromagnetic phase (1) and pure anti-ferromagnetic phase (2) are constructed from a single causal state, while (4) needs two causal states. The ordered phase (3) requires tree causal state and it only happens for $B\neq 0$. 

\subsection{Persistent biased one dimensional random walk}

\begin{figure}[!t]
\centering
\includegraphics[scale=2]{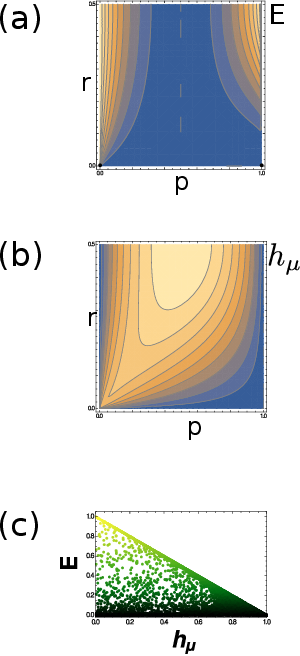}
\caption{Excess entropy (a) and entropy density (b) for different values of the probability of a right $r$ movement and the probability of persistent $p$ movement. (c) The complexity-entropy diagram of the persistent random walker, $10^5$ points with random parameters $(r,p)$ between 0 and 1 were used. 
}\label{fig:pbrw}
\end{figure}

\begin{figure}[!t]
\centering
\includegraphics[scale=1]{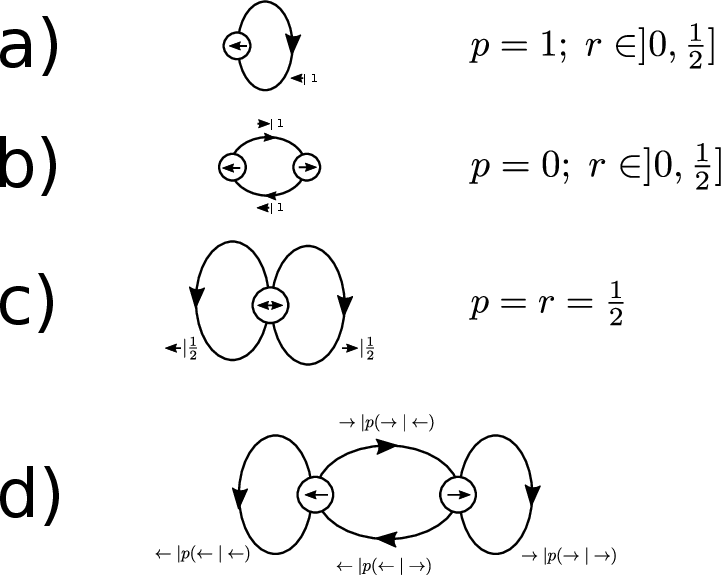}
\caption{$\epsilon$-machine diagrams as a finite state machine for the persistent random walker at different values of the ``right'' movement probability $r$ and the persistence probability  $p$.
}\label{fig:fsm}
\end{figure}

Consider a one dimensional random walk in which at each event the walker moves one step in the same direction as the previous event with probability $p$ (the probability of changing the direction of movement is then $1-p$), we also define the probability that the random walker will move to the right (left) at each event by $r$ ($1-r$). The described model is known as the persistent biased random walk\cite{pottier96,pelayo07} and has been found to be isomorphous with the nearest neighbor Ising chain. The developed formalism has been used to calculate the excess entropy as a function of the probabilities $r$ and $p$, as well as the entropy density, both are shown in figure \ref{fig:pbrw}.  The $\epsilon$-machine reconstruction for different value range of $r$ and $p$ is shown in figure \ref{fig:fsm}. There should be a symmetry in $r$ and $1-r$, as there is no distinction between ``right'' and ``left'', and indeed this is the case, so only values of $r$ up to $1/2$ are shown.  For arbitrary values of the probabilities the finite state machines has two causal states (figure \ref{fig:fsm}d) and disorder is a consequence of the competing ``interaction'' between (anti)persistence of motion direction whose strength is given by $p$, and the movement in some fixed direction measured by $r$. 

If $r$ is small, and $p$ is near one, then persistence takes charge and the random walk will be mostly to the left with seldom changes of direction, this is seen in the low values of $h_{\mu}$. At $p=1$, persistence overtakes any other behavior, and the $\epsilon$-machine is a single causal state equivalent to a ferromagnetic state (figure \ref{fig:fsm}a). If $r$ is small ($\approx0$) and $p$ is also small ($\approx 0$), the random walker has a strong tendency to change the movement direction at every step, which is the absolute dominant behavior at $p=0$ resulting in a perfect anti-ferromagnetic state (figure \ref{fig:fsm}b). At $p=r=1/2$ the maximum state of disorder is reached which is equivalent to an unbiased coin tossing (figure \ref{fig:fsm}c). In the entropy plot (Fig \ref{fig:pbrw}a) a large central region of almost no structure is seen, this same region is where large value of disorder is achieved as seen in the $h_{\mu}$ plot (Fig. \ref{fig:pbrw}b). 

Finally, a complexity-entropy map\cite{feldman08} of the model is shown in figure \ref{fig:pbrw}c. Compare the map with that of the $1/2$-nearest neighbor Ising model. 

\section{Conclusion}

$\epsilon$-machines reconstruction or computational mechanics, is a powerful tool in the analysis of complexity, which has been used in a wealth of different theoretical and practical situations. Unfortunately, for the case of the one dimensional Ising model, previous treatment were flawed by improper treatment of the model as a Markov random field, which, it gave wrong numerical results and misleading behaviors for specific realization of the model. In this paper we developed the complete formalism for $\epsilon$-machine construction of the one dimensional Ising model. The given  expressions can be then used to model specific instance of the interaction Hamiltonian, opening the way to the correct use of computational mechanics for the analysis of complexity behavior in the one-dimensional Ising model. A computer library with the full implementation of the framework can be requested from the authors. 

\begin{acknowledgments}
This work was partially financed by FAPEMIG under the project BPV-00047-13. EER which to thank PVE/CAPES for financial support under the grant 1149-14-8. Infrastructure support was given under project FAPEMIG APQ-02256-12.
\end{acknowledgments}


\end{document}